\documentclass[3p,times,twocolumn]{elsarticle}

\usepackage{ecrc}

\volume{00}

\firstpage{1}

\journalname{Nuclear Physics B Proceedings Supplement}

\runauth{R.~Nevzorov, S.~Pakvasa}

\jid{nuphbp}

\jnltitlelogo{Nuclear Physics B Proceedings Supplement}

\usepackage{amssymb}
\usepackage[figuresright]{rotating}

\begin{document}

\begin{frontmatter}

%% Title, authors and addresses

%% use the tnoteref command within \title for footnotes;
%% use the tnotetext command for the associated footnote;
%% use the fnref command within \author or \address for footnotes;
%% use the fntext command for the associated footnote;
%% use the corref command within \author for corresponding author footnotes;
%% use the cortext command for the associated footnote;
%% use the ead command for the email address,
%% and the form \ead[url] for the home page:
%%
%% \title{Title\tnoteref{label1}}
%% \tnotetext[label1]{}
%% \author{Name\corref{cor1}\fnref{label2}}
%% \ead{email address}
%% \ead[url]{home page}
%% \fntext[label2]{}
%% \cortext[cor1]{}
%% \address{Address\fnref{label3}}
%% \fntext[label3]{}

\dochead{}
%% Use \dochead if there is an article header, e.g. \dochead{Short communication}

\title{Nonstandard Higgs decays in the $E_6$ inspired SUSY models}

%% use optional labels to link authors explicitly to addresses:
%% \author[label1,label2]{<author name>}
%% \address[label1]{<address>}
%% \address[label2]{<address>}

\author[UoA,ITEP]{R.~Nevzorov}

\author[UH]{S.~Pakvasa}

\address[UoA]{ARC Centre of Excellence for Particle Physics at the Tera--scale,\\
School of Chemistry and Physics, University of Adelaide, Adelaide SA 5005, Australia}
\address[ITEP]{Institute for Theoretical and Experimental Physics, Moscow, 117218, Russia}
\address[UH]{Department of Physics and Astronomy, University of Hawaii, Honolulu, Hawaii 96822, USA }

\begin{abstract}
%\noindent
We consider the exotic decays of the SM-like Higgs state within the $E_6$ inspired
supersymmetric (SUSY) models. In these models the lightest SUSY particle
(LSP) can be substantially lighter than $1\,\mbox{eV}$ forming hot dark
matter in the Universe. The next--to--lightest SUSY particle (NLSP) also tend
to be light. We argue that the NLSP with GeV scale mass may lead to the substantial
branching ratio of the nonstandard decays of the SM--like Higgs boson into NLSPs.
\end{abstract}

\begin{keyword}
%% keywords here, in the form: keyword \sep keyword
Higgs boson \sep supersymmetric models \sep Grand Unified Theories

%% MSC codes here, in the form: \MSC code \sep code
%% or \MSC[2008] code \sep code (2000 is the default)
\PACS 14.80.Da \sep 12.60.Jv \sep 12.60.Cn
\end{keyword}

\end{frontmatter}

%%
%% Start line numbering here if you want
%%
% \linenumbers

%% main text
\section{Introduction}
\label{intro}

The discovery of Higgs boson may provide a window into new physics beyond
the Standard Model (SM). Indeed, physics beyond the SM may affect the Higgs
decay rates to SM particles and give rise to new channels of Higgs decays.
In this context it is especially interesting to consider the nonstandard decays
of the SM--like Higgs boson within well motivated extensions of the SM.
Here we focus on the $E_6$ inspired SUSY models which are based on the
$SU(3)_C\times SU(2)_W\times U(1)_Y\times U(1)_{N}$ gauge group,
where
\begin{equation}
U(1)_N=\frac{1}{4} U(1)_{\chi}+\frac{\sqrt{15}}{4} U(1)_{\psi}\,.
\label{1}
\end{equation}
The two anomaly-free $U(1)_{\psi}$\quad and $U(1)_{\chi}$
symmetries can originate from the breakings
$E_6\to$ $SO(10)\times U(1)_{\psi}$, $SO(10)\to SU(5)\times U(1)_{\chi}$.
To ensure anomaly cancellation the particle spectrum in these models
is extended to fill out three complete 27-dimensional representations
of the gauge group $E_6$. Each $27$-plet contains one generation of ordinary matter;
singlet fields, $S_i$; up and down type Higgs doublets, $H^{u}_{i}$ and $H^{d}_{i}$;
charged $\pm 1/3$ coloured exotics $D_i$, $\bar{D}_i$. The presence of exotic
matter in $E_6$ inspired SUSY models generically lead to non--diagonal flavour
transitions and rapid proton decay. To suppress flavour changing processes as well
as baryon and lepton number violating operators one can impose a set of discrete
symmetries \cite{King:2005jy, King:2005my}.
The $E_6$ inspired SUSY models with extra $U(1)_{N}$ gauge symmetry and
suppressed flavor-changing transitions, as well as baryon number violating
operators allow exotic matter to survive down to the TeV scale that may lead
to spectacular new physics signals at the LHC which were analysed in
\cite{King:2005jy, King:2005my, King:2006vu, King:2006rh, Athron:2010zz, Athron:2011ew}.
Only in this Exceptional Supersymmetric Standard Model (E$_6$SSM) \cite{King:2005jy, King:2005my}
right--handed neutrinos do not participate in the gauge interactions so that
they may be superheavy, shedding light on the origin of the mass hierarchy in the
lepton sector and providing a mechanism for the generation of the baryon asymmetry
in the Universe via leptogenesis \cite{King:2008qb}.
Recently the particle spectrum and collider signatures associated with it were
studied within the constrained version of the E$_6$SSM
\cite{Athron:2008np, Athron:2009ue, Athron:2009bs, Athron:2011wu,Athron:2012sq}.

\section{$E_6$~ inspired~ SUSY~ models~ with~ exact~ $\tilde{Z}^{H}_2$~ symmetry}
\label{sec-2}

Here we study the nonstandard Higgs decays within the $E_6$ inspired SUSY
models in which a single discrete $\tilde{Z}^{H}_2$ symmetry forbids tree-level flavor-changing
transitions and the most dangerous baryon and lepton number violating operators \cite{Nevzorov:2012hs}.
These models imply that near some high energy scale (scale $M_X$) $E_6$ or its subgroup is broken down to
$SU(3)_C\times SU(2)_W\times U(1)_Y\times U(1)_{\psi}\times U(1)_{\chi}\times Z_{2}^{M}$,
where $Z_{2}^{M}=(-1)^{3(B-L)}$ is a matter parity. Below scale $M_X$ the particle content
of the considered models involves three copies of $27_i$--plets and a set of $M_{l}$ and
$\overline{M}_l$ supermultiplets from the incomplete $27'_l$ and $\overline{27'}_l$
representations of $E_6$. All matter superfields, that fill in complete $27_i$--plets, are odd
under $\tilde{Z}^{H}_2$ discrete symmetry while the supermultiplets $\overline{M}_l$ can
be either odd or even. All supermultiplets $M_{l}$ are even under the $\tilde{Z}^{H}_2$
symmetry and therefore can be used for the breakdown of gauge symmetry. In the simplest
case the set of $M_{l}$ includes $H_u$, $H_d$, $S$ and $L_4$, where $L_4$ and
$\overline{L}_4$ are lepton $SU(2)_W$ doublet and anti--doublet supermultiplets that
originate from a pair of additional $27'_{L}$ and $\overline{27'}_L$.

At low energies (i.e. TeV scale) the superfields $H_u$, $H_d$ and $S$ play the role of
Higgs fields. The vacuum expectation values (VEVs) of these superfields ($\langle H_d \rangle = v_1/\sqrt{2}$,
$\langle H_u \rangle = v_2/\sqrt{2}$ and $\langle S \rangle = s/\sqrt{2}$) break the
$SU(2)_W\times U(1)_Y\times U(1)_{N}$ gauge symmetry down to $U(1)_{em}$ associated
with the electromagnetism. In the simplest scenario $\overline{H}_u$, $\overline{H}_d$
and $\overline{S}$ are odd under the $\tilde{Z}^{H}_2$ symmetry. As a consequence
$\overline{H}_u$, $\overline{H}_d$ and $\overline{S}$ from the $\overline{27'}_l$
get combined with the superposition of the corresponding components from $27_i$ so that the
resulting vectorlike states gain masses of order of $M_X$. On the other hand $L_4$ and
$\overline{L}_4$ are even under the $\tilde{Z}^{H}_2$ symmetry. These supermultiplets
form TeV scale vectorlike states to render the lightest exotic quark unstable. In this
simplest scenario the exotic quarks are leptoquarks.

The $\tilde{Z}^{H}_2$ symmetry allows the Yukawa interactions in the superpotential that
originate from $27'_l \times 27'_m \times 27'_n$ and $27'_l \times 27_i \times 27_k$.
One can easily check that the corresponding set of operators does not contain any
that lead to the rapid proton decay. Since the set of multiplets $M_{l}$ contains only one
pair of doublets $H_d$ and $H_u$ the $\tilde{Z}^{H}_2$ symmetry also forbids unwanted
FCNC processes at the tree level. The gauge group and field content of the $E_6$ inspired
SUSY models considered here can originate from the orbifold GUT models in which the splitting
of GUT multiplets can be naturally achieved \cite{Nevzorov:2012hs}.

In the simplest scenario discussed above extra matter beyond the minimal supersymmetric
standard model (MSSM) fill in complete $SU(5)$ representations. As a result the gauge
coupling unification remains almost exact in the one--loop approximation. It was also
shown that in the two--loop approximation the unification of the gauge couplings in the
considered scenario can be achieved for any phenomenologically acceptable value of
$\alpha_3(M_Z)$, consistent with the central measured low energy value \cite{King:2007uj}.

As mentioned before, the gauge symmetry in the $E_6$ inspired SUSY models being
considered here, is broken so that the low--energy effective Lagrangian of these models is
invariant under both $Z_{2}^{M}$ and $\tilde{Z}^{H}_2$ symmetries. Since
$\tilde{Z}^{H}_2 = Z_{2}^{M}\times Z_{2}^{E}$  the $Z_{2}^{E}$ symmetry associated
with exotic states is also conserved. The invariance of the Lagrangian under the $Z_{2}^{E}$
symmetry implies that the lightest exotic state, which is odd under this symmetry, must be
stable. Using the method proposed in \cite{Hesselbach:2007te, Hesselbach:2007ta, Hesselbach:2008vt}
it was argued that there are theoretical upper bounds on the masses of the lightest and second
lightest inert neutralino states \cite{Hall:2010ix, Hall:2010ny, Hall:2011au, Hall:2013bua}\footnote{We
use the terminology ``inert Higgs'' to denote Higgs--like doublets and SM singlets that do not develop
VEVs. The fermionic components of these supermultiplets form inert neutralino and chargino
states.}. These states are predominantly the fermion components of the two SM singlet
superfields $S_i$ from $27_i$ which are odd under the $Z_{2}^{E}$ symmetry.
Their masses do not exceed $60-65\,\mbox{GeV}$ so that the lightest and second
lightest inert neutralino states ($\tilde{H}^0_1$ and $\tilde{H}^0_2$) tend to be the
lightest exotic particles in the spectrum \cite{Hall:2010ix, Hall:2010ny, Hall:2011au, Hall:2013bua}.

The $Z_{2}^{M}$ symmetry conservation ensures that $R$--parity is also conserved.
Since the lightest inert neutralino $\tilde{H}^0_1$ is also the lightest $R$--parity
odd state either the lightest $R$--parity even exotic state or the lightest $R$--parity
odd state with $Z_{2}^{E}=+1$ must be absolutely stable. In the considered $E_6$
inspired SUSY models most commonly the second stable state is the lightest ordinary
neutralino $\chi_1^0$ ($Z_{2}^{E}=+1$). Both stable states are natural dark matter
candidates.

When $|m_{\tilde{H}^0_{1}}|\ll M_Z$ the couplings of the lightest inert neutralino
to the gauge bosons, Higgs states, quarks and leptons are very small resulting in
very small annihilation cross section for $\tilde{H}^0_1\tilde{H}^0_1\to \mbox{SM particles}$,
making the cold dark matter density much larger than its measured value. In principle,
$\tilde{H}^0_1$ could account for all or some of the observed cold dark matter density if it
had a mass close to half the $Z$ mass. In this case the lightest inert neutralino states
annihilate mainly through an $s$--channel $Z$--boson \cite{Hall:2010ix}, \cite{Hall:2009aj}.
However the usual SM-like Higgs boson decays more than 95\% of the time into either $\tilde{H}^0_1$
or $\tilde{H}^0_2$ in these cases while the total branching ratio into SM particles is suppressed.
Because of this the corresponding scenarios are basically ruled out nowadays \cite{Hall:2013bua}.

The simplest phenomenologically viable scenarios imply that the lightest inert neutralinos
are extremely light. For example, these states can be substantially lighter than
$1\,\mbox{eV}$\footnote{The presence of very light neutral fermions in the particle spectrum
might have interesting implications for the neutrino physics (see, for example \cite{Frere:1996gb}).}.
In this case, light $\tilde{H}^0_1$ forms hot dark matter in the Universe but gives only a very minor contribution
to the dark matter density while the lightest ordinary neutralino may account for all or some of
the observed cold dark matter density.

\section{Exotic Higgs decays}
\label{sec-3}

As discussed earlier, the $E_6$ inspired SUSY models considered here involves three families of
up and down type Higgs--like doublet supermultiplets ($H^{u}_{i}$ and $H^{d}_{i}$) and three
SM singlet superfields ($S_i$) that carry $U(1)_{N}$ charges. One family of the Higgs--like
doublets and one SM singlet develop VEVs breaking gauge symmetry. The fermionic components of
other Higgs--like and singlet superfields form inert neutralino and chargino states.
The Yukawa interactions of inert Higgs superfields are described by the superpotential
\begin{equation}
\begin{array}{c}
W_{IH}=\lambda_{\alpha\beta} S (H^d_{\alpha} H^u_{\beta})+
f_{\alpha\beta} S_{\alpha} (H_d H^u_{\beta})\\[2mm]
+\tilde{f}_{\alpha\beta} S_{\alpha} (H^d_{\beta} H_u)\,,
\end{array}
\label{2}
\end{equation}
where $\alpha,\beta=1,2$\,. Without loss of generality it is always possible to choose the basis
so that $\lambda_{\alpha\beta}=\lambda_{\alpha\alpha}\,\delta_{\alpha\beta}$. In this basis the
masses of inert charginos are given by
\begin{equation}
m_{\tilde{H}^{\pm}_{\alpha}}=\frac{\lambda_{\alpha\alpha}}{\sqrt{2}}\,s\,.
\label{3}
\end{equation}

In our analysis here we choose the VEV of the SM singlet field $s$
to be large enough ($s\simeq 12\,\mbox{TeV}$) to ensure that the experimental
constraints on $Z'$ boson mass and $Z-Z'$ mixing are satisfied.
To avoid the LEP lower limit on the masses of inert charginos we also choose
the Yukawa couplings $\lambda_{\alpha\alpha}$ so that all inert
chargino states have masses which are larger than $100\,\mbox{GeV}$.
In the following analysis we also require the validity of perturbation theory up
to the GUT scale that constrains the allowed range of all Yukawa couplings.

Here we restrict our consideration to the part of the parameter space that corresponds
to $\lambda_{\alpha\alpha} s\gg f_{\alpha\beta} v,\, \tilde{f}_{\alpha\beta} v$.
In that limit two lightest inert neutralino states $\tilde{H}^0_{1}$ and $\tilde{H}^0_{2}$ are
predominantly inert singlinos. These states tend to be substantially lighter than $100\,\mbox{GeV}$.

When the SUSY breaking scale $M_S$ is considerably larger than the electroweak (EW) scale, the mass matrix
of the CP--even Higgs sector has a hierarchical structure and can be diagonalized using the perturbation
theory \cite{Kovalenko:1998dc, Nevzorov:2000uv, Nevzorov:2001um, Nevzorov:2004ge, Miller:2003ay}.
Here we are going to focus on the scenarios with moderate values of $\tan\beta$ ($\tan\beta< 2-3$).
For these values of $\tan\beta$ the mass of the lightest CP--even Higgs boson $m_{h_1}$ is very sensitive
to the choice of the coupling $\lambda(M_t)$. In particular, in order to get $m_{h_1}\simeq 125\,\mbox{GeV}$
the coupling $\lambda(M_t)$ must be larger than $g'_1\simeq 0.47$. When $\lambda\gtrsim g'_1$, the
qualitative pattern of the Higgs spectrum is rather similar to the one which arises in the PQ symmetric
NMSSM \cite{Nevzorov:2004ge, Miller:2003ay, Miller:2003hm, Miller:2005qua, Panagiotakopoulos:2000wp}.
In the considered limit the heaviest CP--even, CP--odd and charged states are almost degenerate and lie beyond
the $\mbox{TeV}$ range while the mass of the second lightest CP--even Higgs state is set by $M_{Z'}$
\cite{King:2005jy}. In this case the lightest CP--even Higgs boson is the analogue of the SM Higgs field.

In contrast with the MSSM, the lightest Higgs boson in the $E_6$ inspired SUSY models can be heavier than
$110-120\,\mbox{GeV}$ even at tree level. In the two--loop approximation the lightest Higgs boson mass does
not exceed $150-155\,\mbox{GeV}$ \cite{King:2005jy}. Recently, the RG flow of the Yukawa couplings and
the theoretical upper bound on the lightest Higgs boson mass in these models were analysed in the vicinity
of the quasi--fixed point \cite{Nevzorov:2013ixa} that appears as a result of the intersection of the invariant
and quasi--fixed lines \cite{Nevzorov:2001vj}. It was argued that near the quasi--fixed point the upper bound
on the mass of the SM--like Higgs boson is rather close to $125\,\mbox{GeV}$ \cite{Nevzorov:2013ixa}.

The lightest and second lightest inert neutralinos interact with the $Z$--boson and the SM--like Higgs state.
The corresponding part of the Lagrangian, that describes these interactions, can be presented in the following form
\cite{Nevzorov:2013tta}:
\begin{equation}
\begin{array}{c}
\mathcal{L}_{Zh}=\sum_{\alpha,\beta}\frac{M_Z}{2 v}Z_{\mu}
\biggl(\tilde{H}^{0T}_{\alpha}\gamma_{\mu}\gamma_{5}\tilde{H}^0_{\beta}\biggr) R_{Z\alpha\beta}\\[2mm]
+ \sum_{\alpha,\beta} (-1)^{\theta_{\alpha}+\theta_{\beta}} X^{h}_{\alpha\beta} \biggl(\psi^{0T}_{\alpha}
(-i\gamma_{5})^{\theta_{\alpha}+\theta_{\beta}}\psi^0_{\beta}\biggr) h\,,
\end{array}
\label{5}
\end{equation}
where $\alpha,\beta=1,2$. In Eq.~(\ref{5}) $\psi^0_{\alpha}=(-i\gamma_5)^{\theta_{\alpha}}\tilde{H}^0_{\alpha}$ is
the set of inert neutralino eigenstates with positive eigenvalues, while $\theta_{\alpha}$ equals 0 (1) if the eigenvalue
corresponding to $\tilde{H}^0_{\alpha}$ is positive (negative). The inert neutralinos are labeled according to increasing
absolute value of mass, with $\tilde{H}^0_1$ being the lightest inert neutralino.

We further assume that the lightest inert neutralino is substantially lighter than $1\,\mbox{eV}$ so that it gives
only a very minor contribution to the dark matter density. On the other hand we allow the second lightest inert neutralino
state to have mass in the GeV range. Although these states are substantially lighter than $100\,\mbox{GeV}$ their couplings
to the $Z$--boson can be rather small because of the inert singlino admixture in these states. Therefore any possible signal
which these neutralinos could give rise to at former colliders would be extremely suppressed and such states could remain
undetected.

\begin{table}[ht]
\centering
\begin{tabular}{|c||c||c|}
\hline
                              & i       &   ii    \\\hline\hline
$\lambda_{22}$                & -0.03   &     0  \\\hline
$\lambda_{21}$                & 0         &   0.02   \\\hline
$\lambda_{12}$                & 0         &   0.02   \\\hline
$\lambda_{11}$                & 0.03    &   0  \\\hline\hline
$f_{22}$                      & -0.1         &   0.6    \\\hline
$f_{21}$                      & -0.1         &   0.00245 \\\hline
$f_{12}$                      & 0.00001  &   0.00245 \\\hline
$f_{11}$                      & 0.1          &   0.00001 \\\hline\hline
$\tilde{f}_{22}$              & 0.1       &   0.6     \\\hline
$\tilde{f}_{21}$              & 0.1       &   0.002 \\\hline
$\tilde{f}_{12}$              & 0.000011&   0.002  \\\hline
$\tilde{f}_{11}$              & 0.1         &   0.00001 \\\hline\hline
$|m_{\tilde{\chi}^0_1}|$/GeV  &$2.7\cdot 10^{-11}$  & $0.31\cdot 10^{-9}$ \\\hline
$|m_{\tilde{\chi}^0_2}|$/GeV  & 1.09       &   0.319   \\\hline
$|m_{\tilde{\chi}^0_3}|$/GeV  & 254.6     &   169.7  \\\hline
$|m_{\tilde{\chi}^0_4}|$/GeV  & 255.5     &   169.7  \\\hline
$|m_{\tilde{\chi}^0_5}|$/GeV  & 255.8     &   199.1  \\\hline
$|m_{\tilde{\chi}^0_6}|$/GeV  & 256.0      &   199.4  \\\hline\hline
$|m_{\tilde{\chi}^\pm_1}|$/GeV& 254.6    &   169.7  \\\hline
$|m_{\tilde{\chi}^\pm_2}|$/GeV& 254.6    &   169.7  \\\hline\hline
$|R_{Z11}|$                   & 0.0036   &   $1.5\cdot 10^{-7}$ \\\hline
$|R_{Z12}|$                   & 0.0046   &   $1.7\cdot 10^{-4}$ \\\hline
$|R_{Z22}|$                   & 0.0018   &   0.106   \\\hline\hline
$X^{h_1}_{22}$                & 0.0044   &   0.00094 \\\hline
$\mathrm{Br}(h\rightarrow \tilde{\chi}^0_2 \tilde{\chi}^0_2)$& 4.7\%   & 0.22\% \\\hline
$\mathrm{Br}(h\rightarrow b\bar{b})$                         & 56.6\%   & 59.3\% \\\hline
$\Gamma(h\rightarrow \tilde{\chi}^0_2 \tilde{\chi}^0_2)$/MeV & 0.194   & 0.0088 \\\hline
$\Gamma^{tot}$/MeV                                           & 4.15     & 3.962 \\\hline
\end{tabular}
\caption{Benchmark scenarios for $m_{h_1}\approx 125\,\mbox{GeV}$. The branching ratios and decay widths of the
lightest Higgs boson, the masses of inert neutralinos and charginos as well as the couplings of
$\tilde{H}^0_1$ and $\tilde{H}^0_2$ are calculated for $s=12000\,\mbox{GeV}$, $\lambda=0.6$, $\tan\beta=1.5$,
$m_{H^{\pm}}\simeq m_{A}\simeq m_{h_3}\simeq 9497\,\mbox{GeV}$, $m_{h_2}\simeq M_{Z'}\simeq 4450\,\mbox{GeV}$,
$m_Q=m_U=M_S=4000\,\mbox{GeV}$ and $X_t=\sqrt{6} M_S$.}
\end{table}

The couplings of the Higgs states to the inert neutralinos originate from the superpotential (\ref{2}). If all Higgs states
except the lightest one are much heavier than the EW scale then the couplings of the SM--like Higgs boson to the lightest
and second lightest inert neutralinos are determined by their masses \cite{Hall:2010ix}. Since we assumed that the mass
of $\tilde{H}^0_1$ is lighter than $1\,\mbox{eV}$ the couplings of the lightest Higgs boson to $\tilde{H}^0_1\tilde{H}^0_1$
and $\tilde{H}^0_1\tilde{H}^0_2$ are negligibly small and can be ignored in our analysis. Also because of this the experiments
for the direct detection of dark matter do not set any stringent constraints on the masses and couplings of the lightest and
second lightest inert neutralinos. In the considered case the coupling of the SM--like Higgs state to $\tilde{H}^0_2$
is basically proportional to the second lightest inert neutralino mass divided by the VEV, i.e.
$X^{h}_{22}\simeq |m_{\tilde{H}^0_{2}}|/v$ \cite{Hall:2010ix}. This coupling gives rise to the decays of the lightest
Higgs boson into $\tilde{H}^0_2$ pairs with partial widths given by \cite{Nevzorov:2013tta}
\begin{equation}
\Gamma(h_1\to\tilde{H}^0_{2}\tilde{H}^0_{2})=\frac{(X^{h}_{22})^2 m_{h_1}}{4\pi}\biggl(
1-4\frac{|m_{\tilde{H}^0_{2}}|^2}{m^2_{h_1}}\biggr)^{3/2}\,.
\label{6}
\end{equation}

In order to compare the partial widths associated with the exotic decays of the SM-like Higgs state (\ref{6})
with the Higgs decay rates into the SM particles we specify two benchmark points (see Table 1).
For each benchmark scenario we calculate the spectrum of the inert neutralinos, inert charginos
and Higgs bosons as well as their couplings and the branching ratios of the nonstandard decays of the lightest
CP-even Higgs state. We fix $\tan\beta=1.5$ and $\lambda(M_t)=0.6$. As it was mentioned before, such a large
value of $\lambda(M_t)$ allows $m_{h_1}$ to be $125\,\mbox{GeV}$ for moderate $\tan\beta$.
In Table 1 the masses of the heavy Higgs states are computed in the leading one--loop approximation.
In the case of the lightest Higgs boson mass the leading two--loop corrections are taken into account.

From Table 1 it follows that the structure of the Higgs spectrum is extremely hierarchical.
As a result the partial decay widths that correspond to the decays of the lightest CP-even Higgs state
into the SM particles are basically the same as in the SM. Because of this, for the calculation of the Higgs
decay rates into the SM particles we use the results presented in \cite{King:2012is} where these rates
were computed within the SM for different values of the Higgs mass. When $m_{h_1}\simeq 125\,\mbox{GeV}$
the SM-like Higgs state decays predominantly into $b$-quark. In the SM the corresponding branching ratio is
about $60\%$ whereas the branching ratios associated with Higgs decays into $WW$, $ZZ$ and $\gamma\gamma$
are about $20\%$, $2.1\%$ and $0.23\%$ respectively \cite{King:2012is}.
The total decay width of the Higgs boson near 125 GeV is $3.95\,\mbox{MeV}$.

For the calculation of the Higgs decay rates into $\tilde{H}^0_2\tilde{H}^0_2$ we use Eq.~(\ref{6}).
From this equation one can see that the branching ratios of the SM--like Higgs state into the second lightest inert
neutralinos depend rather strongly on the masses of these exotic particles. When $\tilde{H}^0_2$ is relatively heavy,
i.e. $m_{\tilde{H}^0_{2}} \gg m_b(m_{h_1})$, the lightest Higgs boson decays predominantly into $\tilde{H}^0_2\tilde{H}^0_2$
while the branching ratios for decays into SM particles are suppressed. To ensure that the observed signal associated
with the Higgs decays into $\gamma\gamma$ is not too much suppressed we restrict our consideration here to the
GeV scale masses of the second lightest inert neutralino.

The benchmark scenarios (i)-(ii) demonstrate that one can get extremely light
$\tilde{H}^0_1$ with mass $\sim 0.1-0.01\,\mbox{eV}$, relatively light $\tilde{H}^0_2$, that has a mass of the order
of $1-0.1\,\mbox{GeV}$, and a relatively small value of the coupling $R_{Z12}$ that allows the second lightest inert
neutralino to decay within a reasonable time. In these benchmark scenarios the second lightest inert neutralino
decays into the lightest one and a fermion--antifermion pair via virtual $Z$. Since $R_{Z12}$ is relatively
small $\tilde{H}^0_2$ tend to have a long lifetime. If the second lightest inert neutralino state decays during or
after Big Bang Nucleosynthesis (BBN) it may destroy the agreement between the predicted and observed light element
abundances. To preserve the success of the BBN, $\tilde{H}^0_2$ should decay before BBN, i.e. its lifetime $\tau_{\tilde{H}^0_{2}}$
has to be smaller than something like $1\,\mbox{sec}.$ This requirement constrains $|R_{Z12}|$. Indeed, for
$m_{\tilde{H}^0_{2}}=1\,\mbox{GeV}$ the absolute value of the coupling $R_{Z12}$ should be larger than
$1\cdot 10^{-6}$. On the other hand the value of $|R_{Z12}|$ becomes smaller when the mass of the lightest inert
neutralino decreases. Therefore in general sufficiently large fine tuning is needed to ensure that $|R_{Z12}| \gtrsim 10^{-6}$
for sub-eV lightest inert neutralino state. The constraint on $|R_{Z12}|$ becomes much more stringent with decreasing
$m_{\tilde{H}^0_{2}}$ because $\tau_{\tilde{H}^0_{2}}\sim 1/(|R_{Z12}|^2 m_{\tilde{H}^0_{2}}^5)$. As a result, it
is somewhat problematic to satisfy this restriction for $m_{\tilde{H}^0_{2}}\lesssim 100\,\mbox{MeV}$.

The benchmark scenarios (i)-(ii) presented in Table 2 indicate that the branching ratio of the decays of SM--like
Higgs boson into second lightest inert neutralino can vary from $0.2\%$ to $4.7\%$ (i.e. from $0\%$ to $4.7\%$
for practical purposes) when $m_{\tilde{H}^0_{2}}$ changes from $0.3\,\mbox{GeV}$ to $1.1\,\mbox{GeV}$.
For smaller (larger) values of the second lightest inert neutralino masses, the branching ratio associated with these
nonstandard decays of the lightest CP--even Higgs states is even smaller (larger). At the same time the couplings of
$\tilde{H}^0_1$ and $\tilde{H}^0_2$ to the $Z$--boson are so tiny that the lightest and second lightest inert neutralino states
could not be observed before. In particular, their contribution to the $Z$--boson width tend to be rather small. The $Z$--boson
invisible width is characterized by the effective number of neutrino species $N_{\nu}^{eff}$. Its measured value is
$N_{\nu}^{exp}=2.984\pm 0.008$ whereas in the SM $N_{\nu}^{eff}=3$. The contributions of the lightest
and second lightest inert neutralino states to the $Z$--boson width can be parameterized similarly. In the case of benchmark
scenarios (i) and (ii) the effective numbers of neutrino species associated with these contributions are $5.8\cdot 10^{-5}$
and 0.011 respectively.

The second lightest inert neutralino states, that originate from the decays of the SM--like Higgs boson, sequentially decay into
$\tilde{H}^0_1$ and pairs of leptons and quarks via virtual $Z$. Thus, in principle, the exotic decays of the lightest CP--even
Higgs state results in two fermion--antifermion pairs and missing energy in the final state. Nevertheless because coupling
$R_{Z12}$ is quite small $\tilde{H}^0_2$ tend to live longer than $10^{-8}\,\mbox{sec}$. As a consequence the
second lightest inert neutralino state typically decays outside the detectors and will not be observed at the LHC. Therefore the
decay channel $h_1\to\tilde{H}^0_2\tilde{H}^0_2$ normally give rise to an invisible branching ratio of the SM--like Higgs boson.

\section{Conclusions}
\label{sec-4}

In this work we consider the nonstandard Higgs decays within the $E_6$ inspired SUSY models based on the
$SU(3)_C\times SU(2)_W\times U(1)_Y\times U(1)_{N}\times Z_{2}^{M}$ symmetry in which a single discrete
$\tilde{Z}^{H}_2$ symmetry forbids tree-level flavor-changing transitions and the most dangerous baryon
and lepton number violating operators. These models contain at least two states which are absolutely stable and
can contribute to the relic density of dark matter. One of these states is a LSP while another one tends to be the
lightest ordinary neutralino. In the simplest phenomenologically viable scenarios LSP is expected to be substantially
lighter than $1\,\mbox{eV}$ forming hot dark matter in the Universe. At the same time  the lightest ordinary
neutralino can account for all or some of the observed cold dark matter relic density.

The masses of the LSP and NLSP are set by the VEVs of the Higgs doublets. As a consequence they give
rise to nonstandard decays of the lightest Higgs state. Since the couplings of the SM--like Higgs boson to the LSP and NLSP are
determined by their masses LSP does not affect Higgs phenomenology whereas NLSP with GeV scale masses results
in substantial branching ratio of the lightest Higgs decays into NLSPs. After being produced NLSP sequentially
decays into the LSP and pairs of leptons and quarks via virtual $Z$. However due to the small couplings of the LSP
and NLSPs to the $Z$--boson NLSP tends to be longlived particle and decays outside the detectors leading to the
invisible branching ratio of the SM-like Higgs state.

\section{Acknowledgements}
\label{ack}

This work was supported by the University of Adelaide and the Australian Research Council through the ARC
Center of Excellence in Particle Physics at the Terascale.

%% The Appendices part is started with the command \appendix;
%% appendix sections are then done as normal sections
%% \appendix

%% \section{}
%% \label{}

%% References
%%
%% Following citation commands can be used in the body text:
%% Usage of \cite is as follows:
%%   \cite{key}         ==>>  [#]
%%   \cite[chap. 2]{key} ==>> [#, chap. 2]
%%

%% References with BibTeX database:
\nocite{*}
\bibliographystyle{elsarticle-num}
\bibliography{hd-ichep2014}

%% Authors are advised to use a BibTeX database file for their reference list.
%% The provided style file elsarticle-num.bst formats references in the required Procedia style

%% For references without a BibTeX database:

% \begin{thebibliography}{00}

%% \bibitem must have the following form:
%%   \bibitem{key}...
%%

% \bibitem{}

% \end{thebibliography}

\end{document}